\documentclass[nohyper]{JHEP3}

\usepackage{cite}
\usepackage{epsfig}

\newcommand{\be}{\begin{equation}}
\newcommand{\ee}{\end{equation}}
\newcommand{\ba}{\begin{eqnarray}}
\newcommand{\ea}{\end{eqnarray}}

\def\beq{\begin{equation}}
\def\eeq{\end{equation}}
\def\beqa{\begin{eqnarray}}
\def\eeqa{\end{eqnarray}}
\def\eq#1{Eq.~(\ref{#1})}

\def\SDG{\hbox{\tiny SDG}}

\def\DGE{\hbox{\tiny DGE}}

\psfull

\title{The $C$ parameter distribution in $e^+e^-$ annihilation}

\author{Einan Gardi \\
TH Division, CERN, CH--1211 Geneva 23, Switzerland, and \\
Institut f{\"u}r Theoretische Physik, Universit{\"a}t Regensburg,
D--93040 Regensburg, Germany\footnote{present address}}
\author{Lorenzo Magnea \\
Dipartimento di Fisica Teorica, Universit{\`a} di Torino, and\\
INFN, Sezione di Torino, Via P. Giuria, I--10125 Torino, Italy}
  
\abstract{We study perturbative and non--perturbative aspects of the
distribution of the $C$ parameter in $e^+e^-$ annihilation using
renormalon techniques. We perform an exact calculation of the
characteristic function, corresponding to the $C$ parameter
differential cross section for a single off--shell gluon.  We then
concentrate on the two--jet region, derive the Borel representation of
the Sudakov exponent in the large--$\beta_0$ limit and compare the
result to that of the thrust $T$. Analysing the exponent, we
distinguish two ingredients: the jet function, depending on $Q^2 C$,
summarizing the effects of collinear radiation, and a function
describing soft emission at large angles, with momenta of order $Q
C$. The former is the same as for the thrust upon scaling $C$ by $1/6$,
whereas the latter is different.  We verify that the rescaled $C$ distribution
coincides with that of $1 - T$ to next--to--leading logarithmic accuracy,
as predicted by Catani and Webber, and demonstrate that this relation
breaks down beyond this order owing to soft radiation at large angles.
The pattern of power corrections is also similar to that of the thrust:
corrections appear as odd powers of $\Lambda/(Q C)$. Based on the size
of the renormalon ambiguity, however, the shape function is different:
subleading power corrections for the $C$ distribution appear to be
significantly smaller than those for the thrust.}


\preprint{
CERN--TH/2003--129\\
DFTT--6/2003
}

\begin{document}


\section{Introduction}

Event shape distributions have proven to be valuable in pushing
forward our quantitative understanding of jet production and
hadronization and of the interplay between perturbative and
non--perturbative phenomena. Being infrared and collinear safe, these
distributions can be computed order by order in perturbation theory,
without the need to introduce non--perturbative parameters
\cite{ST77,FO78,PA78,DO79,ERT}.  Even at large center--of--mass
energies ($Q$), however, event--shape distributions involve substantial
power corrections. As a consequence, they provide an important tool to
study the onset of non--perturbative physics.

The two--jet region, where the distributions of most event shapes
peak, is particularly important in this regard, and consequently it is
also quite challenging.  First of all, the perturbative analysis
involves large Sudakov logarithms (double logarithms of the event
shape $e$ which vanishes in the limit of two massless jets, examples
being the $C$ parameter, jet masses $\rho_J$, and $t \equiv 1 - T$,
with $T$ the thrust). These logarithms make the perturbative
coefficients of the distributions diverge at any finite order in the
limit $ e \to 0$.  It is only upon resummation that one recovers the
qualitative features of the cross section, namely the fact that it
vanishes in the two--jet limit where radiation is inhibited (Sudakov
suppression). A further complication is that power corrections, which
are associated with large--angle soft emission, are also enhanced in
this limit, as the relevant scale is typically $Q e$.

The resummation of large perturbative
corrections~\cite{CO81,STE86,CSS,CT,CTTW,CTW,CMW,CW,BRO,BSZ} and the
parameterization of power corrections based on renormalon
techniques~\cite{BEN,MW,W94,KS,K95,K98,KOS,KT,DW,NA95,DW97,DMW,AK,DO98,DMS,SMY,BB,BE97,GG99,GA00,EGT,EGJ,G01}
have opened up the way for quantitative predictions in the two--jet
region. While much progress was made during the LEP era, some of the
fundamental questions concerning power corrections have not yet been
fully answered. A classical example is the relation (``universality'')
between corrections to different observables which may be deduced from
renormalon models.  Even a more pragmatic motivation to study event
shapes, namely to provide a precise measurement of the strong
coupling, is hampered by theoretical
uncertainties~\cite{LEP,GG99,EGT,EGJ}. Thus, in spite of having no
active $e^+ e^-$ collider, further theoretical progress is
important. Progress is made nowadays in fixed--order
calculations~\cite{BE02}, as well as in resummation and
parametrization of power corrections~\cite{MA02,FR02,ST03,BA02,DA02,BE03}. Tools
developed in the context of event shapes in $e^+ e^-$ annihilation are
often used in other applications.

One of the aspects where the study of event shapes has taught us
general lessons about QCD is the interplay between perturbative and
non--perturbative corrections. The phenomenological success of
renormalon--based models for power corrections has important
consequences. On the one hand, it shows that perturbative tools are
quite powerful. In a sense, the state--of--the--art theoretical
description of event--shape distributions pushes perturbative
calculations beyond their natural regime of applicability.  In
general, this relies on the understanding that hadronization is a soft
phenomenon, which does not involve significant momentum flow and thus
does not change drastically the perturbative distribution.  On the
other hand, the success of these models implies that there is no way
to quantify non--perturbative corrections and relate them to matrix
elements without controlling first perturbative corrections to all
orders. This stands in sharp contrast to the standard approach taken
when vacuum condensates are estimated in the framework of QCD sum
rules~\cite{SH79}.  The current understanding of power corrections
highlights the significance of running--coupling effects at the
perturbative level.

Our main focus in this paper is the two--jet region. Quantitative
predictions in this region are highly non--trivial, involving large
logarithmic corrections from multiple soft and collinear gluon
radiation, as well as significant running--coupling effects. The
latter have both perturbative and non--perturbative aspects, and some
of the difficulty is related to the ambiguous separation between
them. An important feature of the two--jet region is the fact that, as
$e \to 0$, it becomes necessary to resum not only singular
perturbative contributions, but power corrections as well: in fact for
values of $e = {\cal O} (\Lambda/Q)$, which are not far from the peak
of the distribution in typical LEP data, all power corrections of the
form $(\Lambda/(Q e))^p$ become equally important. The feasibility in
principle of such a resummation was shown in Refs.~\cite{K98,KOS},
extending the factorization of the cross section in the two--jet limit
to power--suppressed effects. Power corrections of this kind can be
summarized in a ``shape function'', which can be naturally combined
with the effects of Sudakov resummation. Renormalon calculations can
then be used to construct highly constrained QCD
models for the shape function.  In this paper, we will essentially
construct such a model for the $C$ parameter distribution, following
the Dressed Gluon Exponentiation (DGE) approach, first developed in
connection with the thrust distribution~\cite{EGT}. In this approach
one computes and resums Sudakov logs and renormalons simultaneously,
obtaining information on leading as well as subleading power
corrections, which can be summarized if desired in an ansatz for the
shape function.  Note that the combination of Sudakov effects with
parametrically enhanced power corrections is definitely not unique to
event--shape distributions. It appears whenever differential cross
sections in QCD are evaluated near a kinematic threshold; other
important examples are Drell--Yan production near the energy
threshold~\cite{KS,BB}, structure functions near the elastic limit
(Bjorken $x$ close to~$1$)~\cite{GKRT,GA02}, and fragmentation
functions of light~\cite{DMS,G01} and heavy quarks~\cite{CA02} near $z
= 1$.  In spite of the different nature of these processes, certain
characteristics of the perturbative expansion are generic~\cite{G01},
and similar techniques are applicable in all cases.

As announced, we will consider here the distribution of a specific
event--shape variable, the $C$~parameter. This variable was first
proposed 25 years ago~\cite{FO78}, and became one of the classical
examples of infrared and collinear safe observables. A
next--to--leading order (NLO) ${\cal O}(\alpha_s^2)$ calculation for
the distribution of the $C$ parameter was performed long ago\cite{ERT}
(the next order is still not available), while the resummation of
Sudakov logs to next--to--leading logarithmic accuracy (NLL) has
become available only recently~\cite{CW}. In the last few years the
example of the $C$ parameter, and in particular its average value, had
an important place in the ongoing debate concerning universality of
power corrections, see {\it e.g.}~\cite{CW,KT,DMS,SMY}. On the other
hand, as far as the resummation of running--coupling effects is
concerned, the $C$ parameter distribution was not analysed, and the
prime example has always been the thrust~\cite{GG99,GA00}.

The study of the thrust and of the heavy--jet mass distributions by
means of DGE~\cite{EGT,EGJ} highlighted several significant features
of higher--order perturbative as well as non--perturbative
contributions. In particular, it was found that subleading Sudakov
logs, that are usually neglected, can give substantial contributions:
they carry the factorial enhancement of renormalons.
Phenomenologically, the refinement of the calculated distribution by
DGE (as compared to the standard resummation to NLL accuracy) turned
out to have a dramatic effect on data fits in the peak region. The
most impressive demonstration of this fact is the agreement between
the non--perturbative parameters extracted from the thrust and the
heavy--jet mass distributions~\cite{EGJ}.  The effect this resummation
has on the extracted value of the coupling is also substantial.  It
is, of course, of great interest to extend the DGE analysis to other
event--shape variables and learn which features are generic and which
depend on the observable. The purpose of this paper is to perform such
an analysis for the $C$ parameter.

We proceed as follows. In the next section we recall the definition of
the $C$ parameter, present the kinematics in a three--particle final
state where the gluon is off shell, and compute the characteristic
function. The final result is summarized in Appendix A.  In Section 3
we expand the characteristic function at small $C$ (details are given
in Appendix B) in order to identify the source of logarithmically
enhanced terms which dominate the two--jet limit; next, we construct a
Borel representation of the Sudakov exponent, and we use it to extract
perturbative as well as non--perturbative information on the
distribution, comparing it with the case of the thrust. Our results
are summarized and briefly discussed in Section 4.

\section{The characteristic function}

The $C$ parameter for electron--positron annihilation events was
originally\footnote{Another definition has been introduced
in~\cite{FO78}; see also \cite{ERT}.}
defined~\cite{PA78,DO79} as
\be
C = 3 (\lambda_1 \lambda_2 + \lambda_2 \lambda_3 + 
  \lambda_3 \lambda_1)~, 
\ee
where $\lambda_{\alpha}$ are the eigenvalues of the matrix
\be 
\Theta_{\alpha \beta} = \frac{1}{\sum_j \left\vert 
{\bf p}^{(j)} \right\vert } \, \sum_i \frac{{\bf p}_{\alpha}^{(i)} \,
{\bf p}_{\beta}^{(i)}}{\left| {\bf p}^{(i)} \right|}~, 
\ee 
and ${\bf p}_{\alpha}^{(i)}$ are the spatial components ($\alpha = 1,
2, 3$) of the $i$--th particle momentum in the center--of--mass
frame. The sums over $i$ and $j$ run over all the final state
particles.
 
A related definition in terms of Lorentz invariants is
\be
C = 3 - \frac32 \sum_{i, j} \frac{(p^{(i)} \cdot p^{(j)})^2}{(p^{(i)}
\cdot q) \, ( p^{(j)} \cdot q)}~,
\label{C_def}
\ee
where $p^{(i)}$ is the four--momentum of the $i$--th particle and $q$
is the total four momentum, $q = \sum_i p^{(i)}$.  The two definitions
are equivalent provided particle masses are neglected (for a discussion
of mass effects in power corrections, see~\cite{SW}).

The $C$ parameter varies in the range $0 \leq C \leq 1$. $C = 0$
corresponds to a perfect two--jet event (with massless jets), while $C
= 1$ characterizes a spherical event. Planar events, including, in
particular, the ${\cal O}(\alpha_s)$ perturbative result, are
distributed in the range $0 \leq C \leq 3/4$.

The information needed to perform renormalon resummation at the level
of a single dressed gluon (the large--$\beta_0$ limit) and,
eventually, learn about power corrections, is present in the
leading--order differential cross section, calculated with an
off--shell gluon~\cite{BEN,DMW,BBB}, sometimes called ``characteristic
function''.  In this section we present an exact calculation of the
characteristic function for the distribution of the $C$ parameter. An
analogous calculation for the thrust was performed
in~\cite{GA00}. Although the calculation and the general structure of
the result are similar, in the case under consideration the
expressions involved are significantly more complicated.

It should be noted that the renormalon calculation we perform treats
the decay of the gluon inclusively. The characteristic function
depends only on the total invariant mass of the particles eventually
produced by the gluon and not on their separate momenta.  Since
event--shape variables such as the $C$ parameter are sensitive to the momenta
of individual final--state particles, our calculation differs, for
example, from a strict large--$N_f$ limit. This point was first noted
in~\cite{NA95}, and was since addressed in various occasions, first in
applications to single--particle inclusive cross sections~\cite{BE97},
then in the context of the conjectured universality of power
corrections~\cite{DO98}. In the thrust case the inclusive
approximation is good, in the sense that higher--order perturbative
terms are numerically close to the strict large--$N_f$
result~\cite{GG99,GA00,EGT} (larger differences occur if one considers
purely non--Abelian contributions). The approximation is especially
good in the two--jet limit in which we are interested.  In the case of
the average $C$ parameter, Smye~\cite{SMY} has performed a detailed
analysis of the effect non--inclusive contributions have on the
coefficient of the $1/Q$ correction, within the framework
of~\cite{DO98}. Also in this case the inclusive approximation is close
to the large--$N_f$ result, while larger differences appear in the
non--Abelian part. An analysis of this kind has not yet been performed
for the distribution. We observe also that it is unclear to what
extent non--inclusive contributions can consistently be treated within
the framework of the dispersive approach, since they contain terms
that are genuinely unrelated to running--coupling effects, even in the
abelian limit.

It should be kept in mind that, with present technology, renormalon
models can only be used as a tool to gather information on the general
structure of power corrections, and possibly a rough estimate on their
size. For such purposes, we believe that the inclusive approximation
is sufficient, a view which is supported by the phenomenological
success of the data fits performed with it, and also by the small
relative size of the non--inclusive correction in the cases in which
it was computed. Given the complexity of the calculation in the case
of the $C$ parameter, we will not attempt here a calculation in the
strict large--$N_f$ limit. A calculation of the distribution along the
lines of \cite{SMY} would in any case be welcome, since it would
improve the control on phase--space effects, and gauge the stability of
our result.

In order to perform the calculation of the characteristic function we
first calculate the $C$ parameter for a three--particle final state: a
quark and an antiquark with momenta $p_1$ and $p_2$ ($p_1^2 = p_2^2 =
0$) and an off--shell gluon with momentum $k$. It is convenient to
define the normalized gluon virtuality by $\xi \equiv k^2/q^2$, and
the variables
\ba 
x_1 & = & 2 p_1 \cdot q/q^2~, \nonumber \\
x_2 & = & 2 p_2 \cdot q/q^2~, \label{ourv} \\ 
y & = & 2 k \cdot q/q^2 \, = \, 2 - x_1 - x_2~, \nonumber 
\ea
which correspond to the energy fractions in the centre--of--mass frame.
Using these variables one finds
\be 
{\bf c}(x_1, x_2, \xi) \, \equiv \, C/6 \, = \,
\frac{(1 - x_1)(1 - x_2)(1 - y + 2 \xi) - \xi^2}{x_1 x_2 y}~.
\label{C}
\ee
Here we rescaled the shape variable by a factor of 6; we will compute the
differential cross section for the rescaled variable $c$, from which
the standard observable can be readily obtained, as
\be
\frac{1}{\sigma} \frac{d \sigma}{d C}(C, Q^2) = \frac16 \left.
\frac{1}{\sigma} \frac{d \sigma}{d c}(c, Q^2) \right\vert_{c = C/6}~.
\label{risc}
\ee
Depending on the precise interpretation of the definition of the $C$
parameter for massive particles there can be in \eq{C} an additional
term (from $i$ and $j$ in (\ref{C_def}) both corresponding to the
gluon) of the form $ - \xi^2/y^2$. Here, however, we will be using
\eq{C} as written. As we shall see below, in the small--$c$ region, $c
= {\cal O}(\xi/y)$, so this term is negligible.

The renormalon--resummed differential cross section in the single
dressed gluon approximation is
\be
\frac{1}{\sigma} \frac{d \sigma}{d c} (c, Q^2) = - \frac{C_F}{2 \beta_0}
\, \int_{0}^{1}{d \xi} \, \frac{d {\cal F}(\xi, c)}{d \xi} \, A(\xi Q^2)~,
\label{diff_c}
\ee
where $\beta_0 = \frac{11}{12} C_A - \frac16 N_f$, and $A(\xi Q^2)$ is the
large--$\beta_0$ running coupling ($A=\beta_0\alpha_s/\pi$) on the 
time--like axis, admitting the Borel representation
\be
A(\xi Q^2) = \int_0^{\infty} d u \, \left( Q^2/\Lambda^2 \right)^{- u}
  \, \frac{\sin \pi u}{\pi u} \, {\rm e}^{\frac53 u} \, \xi^{- u}\,,
\label{axiq}
\ee
where $\Lambda$ is the QCD scale in the ${\overline {\rm MS}}$ scheme,
the constant $5/3$ comes from the renormalization of the fermion loop
in this scheme, and the sine factor originates in the analytic
continuation to the time--like axis.

The characteristic function is of the form
\be
{\cal F}(\xi, c) = \int d x_1 d x_2 \,
{\cal M}(x_1, x_2, \xi) \, \delta \left({\bf c}(x_1, x_2, \xi) - c \right)~.
\label{F}
\ee
Here ${\cal M}$ is the squared matrix element for $\gamma^*
\longrightarrow q \overline{q} g$ (with the coupling and the colour
factor extracted), and is given by
\be
{\cal M}(x_1, x_2, \xi) = \frac{(x_1 + \xi)^2 + (x_2 + \xi)^2}{(1 -
  x_1)(1 - x_2)} - \frac{\xi}{(1 - x_1)^2} - \frac{\xi}{(1 - x_2)^2}~.
\ee
Phase space (illustrated in figure~\ref{contours}) is limited by the
restrictions
\ba
\label{limits}
{\rm hard\, limit} & \hspace*{30pt} & x_1 + x_2 \geq 1 - \xi~, \nonumber \\
{\rm soft\, limit} & \hspace*{30pt} & (1 - x_1)(1 - x_2) \geq \xi~,
\ea
where the hard and soft limits correspond to the upper and lower
bounds on the gluon energy fraction $y = 2 - x_1 - x_2$.

Note that along the soft limit, which will be our main interest in the next
section, the expression for $c$ in \eq{C} simplifies a lot. One gets
\be
\left. {\bf c}(x_1, x_2, \xi) \right\vert_{\rm soft} = \xi/y~.
\label{bou}
\ee
Along this phase--space boundary, the maximal value for $c$ with fixed
$\xi$ is obtained for $x_1 = x_2 = 1 - \sqrt{\xi}$, so $y = y_{\min} =
2 \sqrt{\xi}$ and $c_{\max}^{{\rm soft}} = \sqrt{\xi}/2$. This is the
relevant phase--space limit for large--angle soft gluon emission.  The
minimal value of $c$ is obtained in the intersection with the hard
limit, for either $x_1 = 0$ or $x_2 = 0$, so $y = y_{\max} = 1 + \xi$
and $c_{\min} = \xi/(1 + \xi)$. This is the relevant limit for
collinear gluon emission. The maximum value of $c$ is attained away
from the boundaries of phase space, as seen in Figure 1. It can be
determined by maximizing the explicit expression in \eq{C} within the
physical region, and for small $\xi$ is given by $c_{\max} = 1/8 + 3
\xi/4 + {\cal O} (\xi^2)$.
\FIGURE{
\epsfig{file=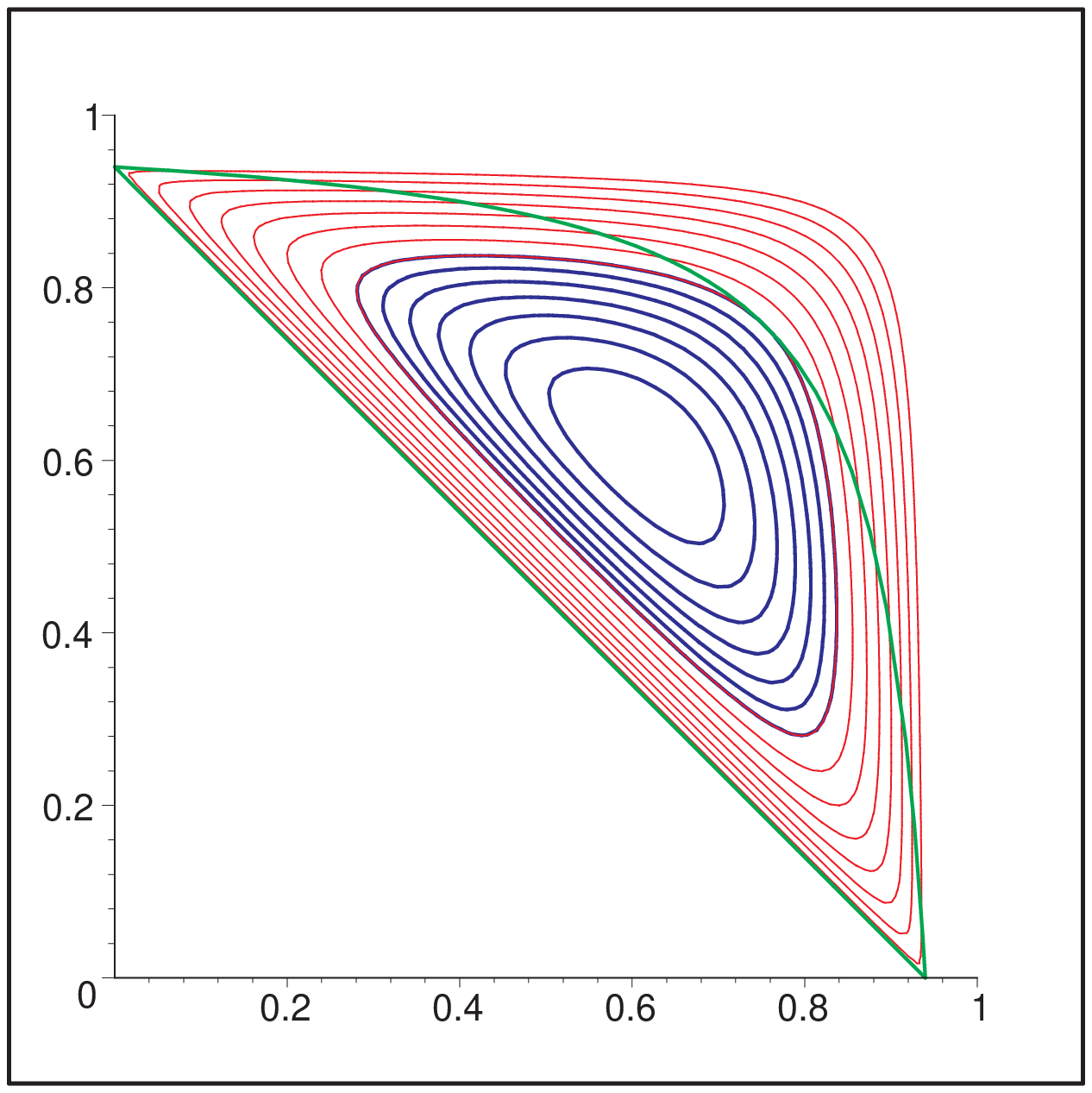,angle=-90,width=12.55cm}
\vspace*{20pt}
\caption{The $x_1$ -- $x_2$ plane for a fixed $\xi=0.06$. The
phase--space limits (\ref{limits}) (green) are shown together with
contours of fixed $c$ belonging to the two regions in (\ref{Fy}):
contours that close within the phase--space boundaries, corresponding
to $F_h$ (thick, blue), and contours that reach the soft phase--space limit,
corresponding to $F_l$ (thin, red).}
\label{contours} }

In order to perform the integrations in~(\ref{F}), it is convenient to
change variables from $x_1$ and $x_2$ into $y$ and $z = (1 - x_2)/y$. In
the new variables the integral has the form
\ba
\label{Fyz}
{\cal F}(\xi, c) & = & \int_{2 \sqrt{\xi}}^{1 + \xi} \frac{dy}{y}
  \int_{\frac12 \left(1 - \sqrt{1 - 4 \xi/y^2} \right)}^{\frac12
  \left(1 + \sqrt{1 - 4 \xi/y^2} \right)} d z
  \, \delta \left((z - z_+)(z - z_-) \frac{y (1 + 2 \xi - y (1 +
  c))}{(1 - y z)(1 - y (1 - z))} \right)
  \nonumber \\ && \nonumber \\ &&
  \,\left[ \frac{(1 - y (1 - z))^2 + (1 - y z)^2 + 2 \xi(2 + \xi - y)}{z
  (1 - z)} - \frac{\xi}{z^2} - \frac{\xi}{(1 - z)^2} \right]\,,
\ea
where
\be
z_{\pm} = \frac12 \left[ 1 \pm \sqrt{1 - \frac{4 (c y (1 - y) +
  \xi^2)}{y^2 (1 + 2 \xi - y (1 + c))}} \right]\,.
\label{Zpm}
\ee
Here the symmetry $x_1 \longleftrightarrow x_2$ appears as $z
\longleftrightarrow 1-z$. Using this symmetry, the integral over $z$
equals twice the integral between the lower limit in \eq{Fyz} and $z =
1/2$, where only $z = z_-$ in the $\delta$ function is relevant.  The
condition $z_- \geq {\frac12 \left(1 - \sqrt{1 - 4\xi/y^2}\right)}$
for physical values of $c$ implies that
\be
\frac{\xi}{c} \, \leq \,y \, \leq \, \frac{1 + 2 \xi}{1 + c} \equiv y_0\,.
\label{z_in_range}
\ee
The condition that $z_-$ be real, on the other hand, implies that
\be
(1 + c) y^3 - (1 + 2 \xi + 4 c) y^2 + 4 c y + 4 \xi^2 \leq 0\,.
\label{cub}
\ee
It turns out that within the physical region all three roots $y_{1,2,3}$
of \eq{cub} are real. Choosing $y_1 \leq y_2 \leq y_3$, \eq{cub} requires
that
\be
y_2 \leq y \leq y_3~.
\label{z_real}
\ee
Investigating the solutions of \eq{cub}, one finds that $y_2 >
{\xi}/c$ for $c > \sqrt{\xi}/2$, while $y_2 \leq {\xi}/c$ for $c \leq
\sqrt{\xi}/2$, so the conditions~(\ref{z_real}) and~(\ref{z_in_range})
together imply that the lower integration limit over $y$ is $y_2$ for
$c> \sqrt{\xi}/2$, but it is $\xi/c$ for $c\leq \sqrt{\xi}/2$. The
result is therefore
\ba
\label{Fy}
{\cal F}(\xi, c) & = & \left\{
\begin{array}{ll}
  \displaystyle{{\cal F}_{l}(\xi, c) \equiv \int_{\xi/c}^{y_3} d y
  f(y, c, \xi)} & \hspace*{40pt} \frac{\xi}{1 + \xi} \, \leq \, c \,
  \leq \, \frac{\sqrt{\xi}}{2}~,
  \\
  \\
  {\cal F}_{h}(\xi, c) \equiv \displaystyle{\int_{y_2}^{y_3} d y
  f(y, c, \xi)} & \hspace*{40pt} \frac{\sqrt{\xi}}{2} \, < \, c \,
  \leq \, c_{\max} = \frac18 + \frac34 \xi + {\cal O}(\xi^2)~,
\end{array}
\right.
\ea
where
\ba
\label{f}
f(y, c, \xi) & = & \frac{2}{(1 + c)^2} \frac{y (1 + \xi - y)^3}
  {(c y (1 - y) + \xi^2)^2 \, \sqrt{(y_0 - y)^3 (y_3 - y)(y -
  y_1)(y - y_2)}} \nonumber \\
  & & \times \Big[
  - c \, (1 + c) \, y^4 + ((5 \, c + 2 \, c^2 + 1) \, \xi + 3 \, c
  + 4 \, c^2) \, y^3 \nonumber \\
  & & \,\,\,\,\,\, + ((- 2 - 4 \, c) \, \xi^2 + (- 2 \, c^2 - 1 - 10 \, c)
  \, \xi - 4 \, c - 4 \, c^2) \, y^2 \\ \nonumber
  & & \,\,\,\,\,\, + ((- 2 \, c - 2) \, \xi^3 - 2 \, \xi^2 + 8 \, \xi \,c
  + 2 \, c) \, y + 4 \, \xi^4 + 8 \, \xi^3 + 2 \, \xi^2 \Big]\,.
\ea
The two regions of the parameter space of \eq{Fy} are distinguished by
whether, for fixed $\xi$ and $c$, the soft phase--space limit in
\eq{limits} can be reached. The corresponding classes of contours are
shown in Figure \ref{contours} in the $x_1$ -- $x_2$ plane. A similar
situation occurs in the case of the thrust distribution~\cite{GA00}.
In both cases the characteristic function is given by two different
analytic function in these two regions.

Finally, ${\cal F}(\xi, c)$ can be expressed as a sum of standard
elliptic integrals. These are complete elliptic integrals for ${\cal
F}_{h}(\xi,c)$ and incomplete ones for ${\cal F}_{l}(\xi,c)$. The
explicit expressions are given in Appendix~A.

\section{The Sudakov exponent}

Having calculated the characteristic function for the distribution of
the $C$ parameter, we can use DGE to compute the Sudakov exponent for
the two--jet limit in the large--$\beta_0$ limit. The procedure we
follow was introduced in~\cite{EGT}, where the exponent for the thrust
distribution was computed. Similar calculations were done since for
heavy--jet mass distribution~\cite{EGJ}, light--quark fragmentation,
deep inelastic structure functions and Drell--Yan production for
$x\longrightarrow 1$~\cite{G01}, and, most recently, the heavy--quark
fragmentation function~\cite{CA02}.

The Borel representation of \eq{diff_c} can be constructed by using
the explicit expression for $A(\xi Q^2)$ given in \eq{axiq} and
changing the order of integration. One obtains
\be
\left. \frac{1}{\sigma} \frac{d \sigma}{d c}(c, Q^2) \right\vert_{\SDG}
  = \frac{C_F}{2 \beta_0} \, \int_0^{\infty} d u \,
  \left( Q^2/\Lambda^2 \right)^{- u} \, B(c, u)\,,
\label{Borel_c}
\ee
where the subscript stands for a Single Dressed Gluon. The Borel
function is
\be
B(c, u) \, = \, - \, \frac{\sin \pi u}{\pi u} \, {\rm e}^{\frac53 u}
  \, \left[ \int_{0}^{4 c^2}{d \xi} \, \frac{d {\cal F}_h (\xi, c)}{d \xi} 
  \, \xi^{- u} \, + \, \int_{4 c^2}^{c/(1 - c)}{d \xi} \, 
  \frac{d {\cal F}_l(\xi, c)}{d \xi} \, \xi^{- u} \right]\,.
\label{Bint}
\ee

While these expressions are completely general, and can be used to
resum the renormalons in the large--$\beta_0$ limit for arbitrary $c$,
we are interested specifically in Sudakov logs, which dominate the
cross section in the two--jet region.  Starting from the full
characteristic function, the first step is to identify the ${\cal
O}(1/c)$ singular terms, namely those terms that upon integration over
the gluon virtuality in \eq{diff_c} or in \eq{Bint} would lead to
logarithmically enhanced contributions in the perturbative expansion
of the distribution.

The identification of the relevant terms was straightforward in
the case of the thrust, but it is much less so here, given the
complexity of the characteristic function, \eq{Fy}. One can expand the
integrand given in~\eq{f} or the closed--form expressions of Appendix A
at small $c$, but this requires of course to specify how $\xi$ behaves in
the limit considered. Given the
limits of integration in \eq{Bint}, there are two natural $c \to 0$
limits to consider: one with fixed $a \equiv \xi/c$, which is relevant
in the collinear region, and the second with fixed $b \equiv \xi/c^2$,
relevant in the large--angle soft emission region.

On general grounds, one expects Sudakov logs to emerge from the soft
boundary of phase space in \eq{limits}. Figure~\ref{contours} shows
that only for ${\cal F}_l$ the soft boundary of phase space is
relevant, so it is only the second term in \eq{Bint} that is expected
to be relevant. We will verify this statement explicitly below. A
similar situation occurs for the thrust distribution~\cite{EGT}.

The details of the small--$c$ expansion of ${\cal F}_l$ at fixed $a$
and at fixed $b$, and of ${\cal F}_h$ at fixed $b$, are summarized in
Appendix B. The final results for the leading ${\cal O}(1/c)$ terms
are
\ba
\label{Fl_fixed_ab}
\left. {\cal F}_{l} (\xi, c) \right\vert_{{\rm fixed} \, a \equiv \xi/c}
  & = & \frac{1}{c} \left[- 4 \ln a - 3 + 2 a + a^2 \right]
  + {\cal O}(c^0) \nonumber \\
  \left. {\cal F}_{l} (\xi, c) \right\vert_{{\rm fixed} \, b\equiv \xi/c^2}
  & = & \frac{1}{c} \left[- 4 \ln b c - 3 - 8 \ln \left(\frac12
  \left(1 + \sqrt{1 - 4/b} \right) \right) \right] + {\cal O}(c^0)\,,
\ea
while
\be
\left. {\cal F}_{h} (\xi, c) \right\vert_{{\rm fixed} \, b \equiv \xi/c^2}
  = \frac{1}{c} \left[- 4 \ln c - 3 \right] + {\cal O}(c^0)\,.
\label{Fh_fixed_b}
\ee
\eq{Fh_fixed_b}) implies that indeed the first term in \eq{Bint}, the
one involving ${\cal F}_h$, is irrelevant for logarithmically enhanced
terms. In order to compute the Sudakov exponent we therefore
concentrate on $F_l$. Using \eq{Fl_fixed_ab} it is straightforward to
write down an expression for $F_l$ that has the correct ${\cal
O}(1/c)$ terms both for fixed $a$ and for fixed $b$. It is given by

\ba
\label{Fl_expand}
\hspace{-6mm}
\left. {\cal F}_{l} (\xi, c) \right\vert_{{\cal O}(1/c)}
  \! = \! \frac{1}{c} \left[- 4 \ln \left( \frac{\xi}{c} \right) - 3 +
  2 \left( \frac{\xi}{c} \right) + \left( \frac{\xi}{c} \right)^2 -
  8 \ln \left(\frac12 \left(1 + \sqrt{1 - 4 c^2/\xi} \right) \right) 
  \right].
\ea
Taking a derivative of \eq{Fl_expand} and substituting into
\eq{Bint} we find
\be
\left. B(c, u) \right\vert_{\rm logs} \, = \, 2{\rm e}^{\frac53 u} \,
  \frac{\sin \pi u}{\pi u} \, \int_{4 c^2}^{c}{d \xi} \, \xi^{- u}
  \left[\frac{2}{\xi c} - \frac{1}{c^2} - \frac{\xi}{c^3} +
  \frac{8 c}{\xi \sqrt{\xi - 4 c^2} \left(\sqrt{\xi} + \sqrt{\xi - 4 c^2} 
  \right)} \right]\,,
\label{Bint_Fl}
\ee
where we simplified the upper limit without affecting the relevant
(log--enhanced) terms. As expected from \eq{Fl_fixed_ab}, only the
first and the last terms in the square brackets in \eq{Bint_Fl}
contribute to the Sudakov exponent owing to the lower (large--angle)
integration limit, while the first three terms, and not the last,
contribute owing to the upper (collinear) limit.  The result of the
integration is
\be
\left. B(c, u) \right\vert_{\rm logs} \, = \, 2 {\rm e}^{\frac53 u} \,
  \frac{\sin \pi u}{\pi u} \, \left[ 4 (2 c)^{ - 1 - 2 u}
  \frac{\sqrt{\pi} \Gamma(u)}{\Gamma(\frac12 + u)} - c^{- 1 - u}
  \left(\frac2u + \frac1{1 - u} + \frac{1}{2 - u} \right) \right]\,.
\label{Bc}
\ee
This simple expression summarizes all log--enhanced contributions to
the $c$ distribution (\ref{Borel_c}), to any order in perturbation
theory, in the large--$\beta_0$ limit. The result is written as a sum
of two ingredients, each originating in a different region of phase
space: the first, where the relevant scale is $4 Q^2 c^2$ (gluon
virtuality $\xi \simeq 4 c^2$) is associated with large--angle soft
emission (it can be traced through the calculation to the region where
$y \simeq 2 \sqrt{\xi}$ and $x_1 \simeq x_2 \simeq 1 - \sqrt{\xi}$);
the second, where the scale is $Q^2 c$ (gluon virtuality $\xi \simeq
c$) is associated with collinear emission (corresponding to $y \simeq
1 + \xi$, and either $x_1 \simeq 0$ or $x_2 \simeq 0$). Note that
infrared safety is guaranteed in \eq{Bc} thanks to the cancellation
between ${\cal O}(1/u)$ terms between the soft and the collinear
ingredients, so each of these ingredients is ill--defined unless a
factorization prescription is introduced. Note also that \eq{Bc} is
free of renormalon singularities, but the Borel integral
(\ref{Borel_c}) does have convergence constraints at small $c$, which
will turn into renormalons in the Sudakov exponent upon performing the
necessary Laplace transform, as outlined below.

For comparison we quote the analogous result for the thrust
distribution~\cite{EGT,G01}.  Defining $t = 1 - T$,
the Borel function in the large--$\beta_0$ limit is
\be
\left. B(t, u) \right\vert_{\rm logs} \, = \, 2 {\rm e}^{\frac53 u} \,
  \frac{\sin \pi u}{\pi u} \, \left[ \frac{2}{u} \, t^{- 1 - 2 u} -
  t^{- 1 - u} \left(\frac2u + \frac1{1 - u} + 
  \frac{1}{2 - u} \right) \right]\,.
\label{Bt}
\ee
The result for the heavy--jet mass~\cite{EGJ} (in this approximation)
is the same as \eq{Bt} up to an overall factor of $1/2$.  The
similarity, as well as the differences, between \eq{Bc} and \eq{Bt}
are easily understood: the collinear ingredient is {\em identical}
(with $c$ replaced by $t$), while the large--angle soft emission
ingredient is different.  In fact, replacing $c$ by $1 - x$, the
collinear ingredient coincides with the large--$\beta_0$
expression~\cite{G01} for the jet function~\cite{STE86,CSS,CT} which
controls the large--$x$ limit in single--particle inclusive
cross--sections in $e^+e^-$ annihilation~\cite{G01,CA02}, as well as
for structure functions near the elastic limit~\cite{GKRT,GA02}. This
object describes the radiation from an unobserved jet under a
restriction on its invariant mass. It therefore appears in each of
these observables.

As noted above, the large--angle soft emission ingredient for the $c$
distribution in \eq{Bc} turns out to be quite different from that of
the thrust, in \eq{Bt}. This is not surprising: different event--shape
variables weigh differently the soft radiation. One immediately
recognizes that the relevant scale of large--angle soft radiation in
the $c$ distribution is $2 Q c$, while for thrust it is $Q t$. As we
shall see below this factor of~$2$ suppresses subleading logarithmic
corrections, as well as subleading power corrections, in the case of
$c$, as compared to $t$.

We now turn to the computation of the Sudakov exponent.
Owing to the additivity property of the $C$ parameter in the small $C$ limit
with respect to multiple emission, and the factorization property of QCD
matrix elements for soft and collinear radiation, logarithmically enhanced
terms in the perturbative expansion exponentiate in Laplace space.
Consequently, the resummed cross section can be written, in full analogy
with the thrust~\cite{CTTW,EGT}, as
\be
\left. \frac{1}{\sigma} \frac{d \sigma}{d c} (c, Q^2)
  \right\vert_{\DGE} = \int_{ k - {\rm i} \infty}^{k + {\rm i} \infty}
  \frac{d \nu}{2 \pi {\rm i}} \, {\rm e}^{\nu c} \,
  \exp \left[ S(\nu, Q^2) \right]\,,
\label{c_DGE}
\ee
where $k$ is to the right of the singularities of the integrand. The
Sudakov exponent is given~by
\be
S(\nu, Q^2) \equiv \left< e^{- \nu c} \right> = \int_0^{\infty} \,
  d \tilde{c} \, \left. \frac{1}{\sigma} \frac{d \sigma}{d
  {\tilde{c}}} (\tilde{c}, Q^2) \right\vert_{\SDG}
  \left({\rm e}^{- \nu \tilde{c}} - 1 \right)\,,
\label{S}
\ee
where the upper limit of integration was extended to infinity ({\it
i.e.} beyond the range where the physical distribution has support) to
comply with the standard definition of the Laplace transform. The
integral is dominated by the Sudakov region, so that singular terms
for $c \to 0$ in \eq{Borel_c} generate logarithmically divergent terms
for $\nu \to \infty$ in $S(\nu, Q^2)$. It should be noted that, as far
as these singular terms are concerned, and to any logarithmic
accuracy, the Laplace transform is equivalent to a Mellin transform,
where $\nu$ is the moment conjugate to $1 - \tilde{c}$.

Next, we compute $S(\nu, Q^2)$ in the large--$\beta_0$ limit using
\eq{Borel_c} and \eq{Bc}. The result is
\be
S(\nu, Q^2) = \frac{C_F}{2 \beta_0} \, \int_0^{\infty} d u \,
  \left(Q^2/\Lambda^2\right)^{- u} \, B_\nu^c (u)\,,
\label{Borel_S}
\ee
with 
\ba
B_\nu^c (u) & = & \int_0^{\infty} d c \, B(c, u) \, 
  \left({\rm e}^{- \nu c} - 1 \right) = \, 2 {\rm e}^{\frac53 u} \,
  \frac{\sin \pi u}{\pi u} \, \left[\Gamma(- 2 u) \left(\nu^{2 u} - 1 
  \right) 2^{1 - 2 u} \frac{\sqrt{\pi} \Gamma(u)}{\Gamma(\frac12 + u)}
  \right.\nonumber \\ 
  & & \hspace*{80pt} - \left. \Gamma(- u) \left({\nu}^{u} - 1 \right)
  \left(\frac2u + \frac1{1 - u} + \frac{1}{2 - u} \right) \right]\,,
\label{Bnu}
\ea
where terms suppressed by powers of $1/\nu$ were neglected.

Contrary to~\eq{Bc}, the Borel function in Laplace space given by
\eq{Bnu} has renormalon singularities. The appearance of infrared
renormalons as a consequence of taking a Laplace or a Mellin transform
is characteristic of differential cross--sections near a kinematic
threshold~\cite{EGT,G01,GA02,CA02}.  In \eq{Bnu} renormalons appear in
both the large--angle soft emission factor and in the jet function. In
the former, they correspond to corrections that scale as odd powers of
$Q/\nu$, while in the latter to ones that scale as the first two
powers of $Q^2/\nu$, {\it i.e.} twist four and twist six, as in
structure functions. The leading power corrections in the Sudakov
region (so long as $Q^2c\gg \Lambda^2$) are of the former type, and
they can be summed up into a shape function, as discussed
in~\cite{KS,K95,K98,KOS,KT,EGT,EGJ}.  We further address this issue
below.

\eq{Bnu}, together with \eq{Borel_S} and \eq{c_DGE}, is our final
result for the $c$ distribution, calculated by DGE. The experience
gained in the case of the thrust and the heavy--jet mass
distributions~\cite{EGT,EGJ} shows that there is a significant
difference between the resummed cross section \`a la DGE and standard
resummation with NLL accuracy.  Owing to the renormalon factorial
growth, the additional terms (subleading logs) which are resummed by
DGE and neglected otherwise are numerically important.  DGE has a
significant impact on the quality of the description of the
distribution in the peak region, on the extracted non--perturbative
parameters, and, in particular, on the consistency of the latter
between the thrust and the heavy--jet mass distributions. It is
therefore important to analyse more event--shape distributions with the
same methodology, and the road is now open to do so for the $C$
parameter.

It should be emphasized that controlling the large--$\beta_0$ terms is
sufficient for generating the entire set of logs up to NLL
accuracy. This can be done~\cite{EGT} by promoting the running
coupling of~\eq{axiq} to two loops, while replacing the constant $5/3$
in the exponent by the full coefficient of the $x \to 1$ singular term
in the NLO splitting function~\cite{CMW}, according to
\be 
\frac53 \, \longrightarrow \, \frac53 + \left(\frac13 - \frac{\pi^2}{12} 
  \right) \frac{C_A}{\beta_0}\,.
\label{subst}
\ee
One should keep in mind, though, that this replacement is sufficient
only to NLL accuracy, while the correct generalization of the Borel
function~(\ref{Bnu}) beyond the large--$\beta_0$ limit is not
known. See~\cite{GA02} for a relevant discussion.

Of course, as it stands, the integral in \eq{Borel_S} is ill defined
due to its renormalons singularities.  It is straightforward, however,
to perform a principal--value regularization either
analytically~\cite{EGT,EGJ} or numerically~\cite{CA02}. The size of
the ambiguity of \eq{Borel_S}, as measured by the residues of the
singularities, can be used as an estimate of the size of the
corresponding power corrections, as discussed below. Let us also
recall that, in order to use the DGE result in practice, one needs to
match it onto the known fixed--order calculation, to take into account
terms which are not logarithmically enhanced. The procedure~\cite{EGT}
is similar to the one used in a standard NLL resummation
program~\cite{CTTW}.

The location of the renormalons in \eq{Bnu} is identical to the case
of the thrust. For easy comparison we quote the analogous expression
derived in that case, where $\nu$ is the Laplace conjugate variable to
$t = 1 - T$. It is given by
\be
\label{Bnu_t}
B_\nu^t (u) = \, 2{\rm e}^{\frac53 u} \, \frac{\sin \pi u}{\pi u}
  \left[\Gamma(- 2 u) \left(\nu^{2 u} - 1 \right) \frac{2}{u}
  - \Gamma(- u) \left(\nu^u - 1\right) \left(\frac2u + \frac1{1 - u}
  + \frac{1}{2 - u} \right) \right]\,.
\label{bnut}
\ee
In order to extract the perturbative coefficients of the exponent
$S(\nu, Q^2)$ in the large--$\beta_0$ limit, it is sufficient to
expand $B_\nu (u)$ in powers of $u$ and replace $u^n$ by $n!
\left(\beta_0 \alpha_s/\pi \right)^{n + 1}$.  Before looking at the
actual coefficients, let us first note that the difference between the
$c$ distribution, \eq{Bnu}, and the thrust distribution, \eq{Bnu_t},
which is all due to the coefficient of $\left(\nu^{2 u} - 1\right)$,
is relatively small.  Taking the ratio between these two coefficients and
expanding in $u$ we get
\be
2^{- 2 u} \frac{\sqrt{\pi} u \Gamma(u)}{\Gamma(\frac12 + u)} = 1 - 
  \frac{\pi^2}{6} u^2 + 2 \zeta_3 u^3 + {\cal O} (u^4) \, .
\label{diff}
\ee
The fact that the ratio is $1$ at leading order (as it must be, in
order to cancel the $1/u$ singularity of the jet function, which is
identical in the two cases) implies that the leading logs ($L^{n + 1}
\alpha_s^n$ where $L \equiv \ln \nu$, for any $n$) are identical for
the two shape variables; the fact that the next term, of order~$u$, is
missing, implies that also the next--to--leading logs ($L^{n}
\alpha_s^n$, for any $n$) are identical. This confirms the prediction
by Catani and Webber~\cite{CW}.  Differences between the two
distributions appear only at the next--to--next--to--leading
logarithmic level.

In order to illustrate the enhancement of subleading
logarithms~\cite{EGT}, as well as the difference between the $c$
distribution and the thrust in this context, we present below an
expansion of \eq{Bnu} and \eq{Bnu_t} to the first few orders. One
finds for the $c$ parameter
\ba
\label{expand_c}
B_{\nu}^{c}\left(u\right)&=&- 2\,L^{2} + 0.691\,L  \nonumber\\
&&\hspace{-6pt}+\, \left( - 2\,L^{3} - 5.297\,L^{2} + 0.095
\,L\right)\,u\\
&&\hspace{-6pt}+\,\left( - 1.167\,L^{4} - 5.527\,L^{3} -
7.911\,L^{2} - 22.709\,L\right)\,u^{2} \nonumber\\
&&\hspace{-6pt}+\, \left( - 0.5\,L^{5} -
3.262\,L^{4}  - 7.943\,L^{3} - 30.058\,L^{2} - 61.322\, L\right)
\,u^{3}
\nonumber\\\nonumber
&&\hspace{-6pt}+\,\left( - 0.172\,L^{6} - 1.405\,L^{5} - 4.639\,L^{4}
 - 22.488\,L^{3} - 67.59\,L^{2} - 135.22\,L\right)\,u^{4}
\ea
while the thrust gives
\ba
\label{expand_t}
B_{\nu}^{\rm thrust}\left(u\right)&=& - 2\,L^{2} + 0.691\,L
\nonumber\\
&& \hspace{-6pt}+\, \left( - 2\,L^{3} - 5.297\,L^{2} - 6.485\,L\right)\,u
\\
&&
\hspace{-6pt}+\, \left( - 1.167\,L^{4} - 5.527\,L^{3} -
14.491\,L^{2} - 31.655\,L\right)\,u^{2} \nonumber\\
&&
\hspace{-6pt}+\, \left( - 0.5\,L^{5} -
3.262\,L^{4}  - 12.329\,L^{3} - 39.003\,L^{2} - 80.940\,
L\right)\,u^{3} \nonumber\\\nonumber
&&\hspace{-6pt}+\,\left( - 0.172\,L^{6} - 1.405\,L^{5} - 6.832\,L^{4}
 - 28.452\,L^{3}  - 87.21\,L^{2} - 175.80\,L\right)\,u^{4}
\ea
As explained above, the leading and next--to--leading logarithms are
the same. Beyond this order the general trend is similar, however the
increase of the coefficients is milder in the case of $c$.

Finally, we return to the issue of power corrections. One can
construct a parametrization of non--perturbative corrections based on
the ambiguity induced by renormalon singularities in $B_\nu (u)$.  The
residue of a pole at $u = m/2$, where $m$ is an odd integer,
multiplies an ambiguous contribution, and thus a non--perturbative
correction of order $\left(\Lambda\nu/Q\right)^m$ in $S(\nu,Q^2)$ is
necessary to compensate for the ambiguity of the perturbative
result. Summing over $m$, these corrections amount to a multiplication
of the perturbative Laplace--space result
$\exp\left[S(\nu,Q^2)\right]$, entering \eq{c_DGE}, by a
non--perturbative shape function~\cite{KS,K95,K98,KOS,KT,EGT,EGJ} of
the variable $\Lambda\nu/Q$. Since these corrections exponentiate, the
leading $m = 1$ correction generates a shift of the perturbative
distribution~\cite{K95,DW97}, while $m = 3$ and higher--order powers
generate smearing.  Although the large--$\beta_0$ renormalon
calculation is not sufficient to determine the actual magnitude of
power corrections, the size of the residues\footnote{Note that the
large--$\beta_0$ limit does not necessarily provide a good estimate of
the residue. Moreover, in the full theory the singularities are
usually not simple poles.} may be taken as a na\"{\i}ve estimate. The
first few residues (times $\pi$) are summarized in
Table~\ref{tab:residues}.
\renewcommand{\arraystretch}{1.3}
\TABLE{
\begin{tabular}{|c|c|c|l|}
  \hline
  Correction & residue ($c$) & residue ($t$) & ratio ($c/t$) \\ 
  \hline
  $\left(\bar{\Lambda} \nu/Q \right)^1$ & $2 \,\pi$ & $8$ 
  & $\frac{\pi}{4} \simeq 0.79$ \\  
  \hline
  $\left(\bar{\Lambda} \nu/Q \right)^3$ & $- \frac{\pi}{72}$ 
  & $ - \frac{4}{27} $ & $ \frac{3 \, \pi}{32} \simeq 0.29$ \\  
  \hline
  $\left(\bar{\Lambda} \nu/Q \right)^5$ & $ \frac{\pi}{12800}$
  & $\frac{1}{375} $ & $\frac{15 \pi}{512} \simeq 0.092$ \\  
  \hline
  $\left(\bar{\Lambda} \nu/Q \right)^7$ & $- \frac{\pi}{3612672} $
  & $- \frac {1}{30870}$ & $\frac{35 \, \pi}{4096} \simeq 0.027$ \\  
  \hline
  $\left(\bar{\Lambda} \nu/Q \right)^9$ & $ \frac{\pi}{1528823808} $
  & $\frac{1}{3674160} $ & $\frac{315 \, \pi}{131072} \simeq 0.0076 $ \\  
  \hline
\end{tabular}
\caption{The size of the residues of renormalon singularities in the
large--$\beta_0$ limit, based on \eq{Bnu} for the $c$ distribution and
on \eq{Bnu_t} for the thrust $t$. The numbers quoted are $\pi$ times
the coefficients of $\left(\bar{\Lambda} \nu/Q\right)^m$, where we
define $\bar{\Lambda}^2 = \Lambda^2 {\rm e}^{5/3}$. We ignore here the
${\cal O}(1)$ factor $C_F/(2 \beta_0)$ in \eq{Borel_S}.}
\label{tab:residues} }
\renewcommand{\arraystretch}{1.}

Assuming that the large--$\beta_0$ residues do provide some hint on
the size of the corrections, from Table \ref{tab:residues} one would
conclude that the non--perturbative shape functions for the $c$
parameter and the thrust distributions must be significantly
different.  From the perspective of this renormalon model, it seems
therefore that the conjecture of~\cite{KT}, that the same function
(with the same parameters) would be appropriate for both distribution,
should be excluded.  In fact, while the powers of $\Lambda \nu/Q$
identifying the relevant moments of this function are the same for
both variables, subleading corrections to the $c$ distribution are
significantly smaller than the corresponding ones for the thrust.
This means that while the shift of the two distributions is of
comparable magnitude (the first line in the table), the extent to
which subleading non--perturbative corrections smear the $c$
distribution is much smaller than for thrust.

\section{Conclusions}

In this paper we performed a renormalon calculation for the
distribution of the $C$ parameter.  We adopted the dispersive approach
and extracted all--order information from the differential cross
section with a single off--shell gluon. The only approximation we made
with respect to the exact, all--order result in the large--$N_f$
limit, was to treat the gluon decay inclusively.

The characteristic function given in Appendix A can be used to improve
existing fixed--order calculations of the distribution and its first
few moments through the resummation of running--coupling
effects~\cite{GG99,GA00}.  We recall that the perturbative
coefficients of the first few moments of event--shape distributions,
and in particular the average, are dominated by running--coupling
effects, so the impact this resummation has on phenomenology is
significant.

Here we concentrated on the distribution in the two--jet region, where
the perturbative expansion is dominated by Sudakov logs. Starting from
the exact characteristic function, we identified the origin of
logarithmically enhanced terms and computed the Sudakov exponent in
the large--$\beta_0$ limit by Dressed Gluon Exponentiation, similarly
to previous calculations for the thrust and the heavy--jet mass
distributions~\cite{EGT,EGJ}. We showed that the all--order result for
the Sudakov exponent, given in \eq{Bnu}, separates into two
ingredients in a natural way: one is the jet function, depending on
$Q^2 c$, where $c \equiv C/6$, and the other is associated with soft
emission at large angles, with momenta of order~$2 Q c$.  The jet
function~\cite{STE86,CSS,CT} describes the radiation from an
unobserved jet under a restriction on its invariant mass. It can be
defined and computed in a process independent way, and it plays a role
in a large class of differential cross sections near partonic
threshold, which include, in addition to event--shape
distributions~\cite{EGT,EGJ}, the coefficient functions for
single--particle inclusive cross sections in $e^+e^-$ annihilation, as
well as deep--inelastic structure functions. The same large--$\beta_0$
result for the jet function was shown to be relevant for all of these
observables~\cite{G01,GKRT,CA02,GA02}.

Throughout the paper we compared the case of the $C$ parameter to that
of the thrust. As noted first by Catani and Webber~\cite{CW}, the two
distributions are closely related: upon scaling $C$ by $1/6$ the
Sudakov exponents of the two variables coincide to NLL accuracy.
Since the jet functions are the same to any logarithmic accuracy,
differences between the two exponents appear only in the large--angle
soft emission ingredient. The large--$\beta_0$ results, \eq{Bnu} and
\eq{Bnu_t}, indicate that such differences appear at the NNLL order.
The universal phenomenon that subleading logs appear with increasing
numerical coefficients owing to infrared renormalons~\cite{EGT} is
realised, of course, also for the $C$ parameter. The increase of the
coefficients, however, is milder than for thrust. The difference is
largest for the NNLL term at order\footnote{Note that
the coefficients at this order coincide with the exact large $N_f$
result -- they are not influenced at all by the inclusive
approximation~\cite{EGT,EGJ}.} ${\cal O}(u)$,
{\it i.e.} ${\cal O}(\alpha_s^2)$, in
\eq{expand_c}, as compared to \eq{expand_t}.

It is natural to expect that the difference between the thrust and the
$c$ distribution, owing to large--angle soft emission, would be
realised also at the non--perturbative level. While the general
pattern of renormalon singularities in the Sudakov exponent is similar
in the two cases, and the parametrization of non--perturbative
corrections as a shift~\cite{K95,DW97} of the perturbative
distribution, or better, through a convolution with a shape
function~\cite{KS,K95,K98,KOS,KT,EGT,EGJ}, is appropriate in both, it
seems that the corrections themselves are different.  This conclusion
is based on the comparison (Table~\ref{tab:residues}) between the
renormalon residues in the two cases: subleading power corrections for
the $c$ distribution are expected to be smaller and thus a shift
should be a better approximation for the $c$ distribution than it is
for thrust.  We emphasize that the residues computed in the
large--$\beta_0$ limit are not necessarily indicative of the actual
size of the corrections, and it remains to be seen whether future
phenomenological analyses will support these conclusions.


\acknowledgments

We would like to thank G.P.~Korchemsky and V.M.~Braun for very useful
discussions. E.G. would like to thank the DFG for financial support.
L.M. would like to thank the CERN Theory Division for
support during part of this work. Work supported by the italian
Ministry of Education, University and Research (MIUR), under contract
2001023713--006.


\newpage

\appendix

\section{Explicit expression of the characteristic function}

The last integration over the gluon energy fraction $y$ in \eq{Fy} can
be explicitly performed, and the result expressed in terms of standard
elliptic functions. This is a consequence of a well--known theorem of
Legendre, stating that any integral of the general form
\beq
{\cal I}(x) = \int d x R \Big[ x, S (x) \Big]~,
\label{leg}
\eeq
where $R$ is a rational function of its arguments, while $S^2(x)$ is a
polynomial of degree $d \leq 4$, can be expressed as a linear
combination of the three basic elliptic integrals, plus the integral
of a rational function. Our integrand, \eq{f}, clearly fulfills the
requirements. To fix our notation, we define the three basic kinds of
(incomplete) elliptic integrals by
\beqa
F \left(\phi, m\right) & \equiv & \int_0^\phi d \theta \frac{1}{\sqrt{1 - 
m \sin^2 \theta}} = \int_0^{\sin \phi} \frac{d x}{\sqrt{(1 - 
x^2)(1 - m x^2)}}~, \nonumber \\
E \left(\phi, m\right) & \equiv & \int_0^\phi d \theta \sqrt{1 - 
m \sin^2 \theta} = \int_0^{\sin \phi} \sqrt{\frac{1 - m x^2}{1 - x^2}}~,
\label{elli} \\
\Pi \left(n, \phi, m\right) & \equiv & \int_0^\phi d \theta
\frac{1}{(1 - n \sin^2 \theta) \sqrt{1 - m \sin^2 \theta}} =
\int_0^{\sin \phi} \frac{d x}{(1 - n x^2) \sqrt{(1 - x^2)(1 - m x^2)}}~.
\nonumber
\eeqa
The corresponding complete elliptic integrals are obtained by setting
$\phi = \pi/2$, according to ${\bf K} (m) \equiv F (\pi/2, m)$, ${\bf
E} (m) \equiv E (\pi/2, m)$, and ${\bf \Pi} (n, m) \equiv \Pi (n,
\pi/2, m)$. The algorithms to reduce the generic integral in \eq{leg}
to a combination of the standard ones given in \eq{elli}, as well as
the analytic properties of elliptic integrals, are described in some
detail in Ref.~\cite{BAT}.

To express the result of the integration over the gluon energy
fraction $y$ in \eq{Fy}, it is convenient to recall the special values
of $y$ that correspond to singularities in the integrand,
\eq{f}. First, there are the three roots of the cubic
equation~(\ref{cub}), which are real in the physical region and which
we ordered according to $y_1 \leq y_2 \leq y_3$. Recall that the $y$
integration specified in~\eq{Fy} extends over the range from $y_2$ to
$y_3$ for ${\cal F}_h$, or part of it, from $a \equiv \xi/c$ to $y_3$,
for ${\cal F}_l$.  Thus these square--root singularities are either
outside the integration region or on its boundaries. An additional
singularity in the integrand appears at $y = y_0 \equiv (1 + 2 \xi)/(1
+ c)$, however it obeys $y_0 \geq y_3$ so it is always outside the
integration region. Finally there are two double poles outside the
integration region at $y = y_4 \equiv (1 - \sqrt{1 + 4 \xi^2/c})/2$
and at $y = y_5 \equiv (1 + \sqrt{1 + 4 \xi^2/c})/2$.

The arguments of the relevant elliptic integrals can be expressed
in terms of simple ratios of these special values of the gluon energy
fraction. One needs
\beq
\chi \left( \xi, c \right)  = \arcsin \left[
  \sqrt{\frac{(y_0 - y_2)(y_3 - a)}{(y_3 - y_2) (y_0 - a)}} \right]~,
\label{chi}
\eeq
\beq
m \left( \xi, c \right)  = \frac{(y_3 - y_2)(y_0 - y_1)}{(y_0 - 
  y_2) (y_3 - y_1)}~,
\label{mpar}
\eeq
\beq
n \left( \xi, c \right)  = \frac{y_3 - y_2}{y_0 - y_2}~,
\label{n0}
\eeq
\beq
n_4 \left( \xi, c \right)  =  \frac{(y_3 - y_2)(y_4 - y_0)}{(y_0 - 
  y_2)(y_4 - y_3)}~,
\label{n4}
\eeq
\beq
n_5 \left( \xi, c \right)  = \frac{(y_3 - y_2)(y_5 - y_0)}{(y_0 - 
  y_2)(y_5 - y_3)}~.
\label{n5}
\eeq
In terms of these rather intricate functions of $\xi$ and $c$, one can
write the explicit expression for ${\cal F}_{l}(\xi,c)$. It is
\beqa
{\cal F}_{l}(\xi,c) & = & r \left(\xi, c \right) + 
  f \left(\xi, c \right) F \Big[ \chi \left(\xi, c \right), 
  m \left(\xi, c \right) \Big] + e \left(\xi, c \right) 
  E \Big[ \chi \left( \xi, c \right), m \left(\xi, c \right) \Big] 
  \nonumber \\ & + &
  p \left(\xi, c \right) \Pi \Big[ n \left(\xi, c \right), 
  \chi \left(\xi, c \right), m \left(\xi, c \right) \Big] +
  p_4 \left(\xi, c \right) \Pi \Big[ n_4 \left(\xi, c \right), 
  \chi \left(\xi, c \right), m \left(\xi, c \right) \Big] 
  \nonumber \\ & + & 
  p_5 \left(\xi, c \right) \Pi \Big[ n_5 \left(\xi, c \right), 
  \chi \left(\xi, c \right), m \left(\xi, c \right) \Big]~,
\label{flfin}
\eeqa
where
\begin{eqnarray}
r \left(\xi, c \right) & = & \frac{\sqrt{\xi^2 - 4 c^2 \xi}}{c^3 \left(1 +
  c \right)^2 \left(c + 4 \xi^2 \right)} \Big[4 \xi^3 + c \, \xi
  \left(1 + 18 \xi + 32 \xi^2 \right) \nonumber \\
  & + & c^2 \left( 4 + 7 \xi + 20 \xi^2 + 48
  \xi^3 \right) + c^3 \left( 4 + 10 \xi + 2 \xi^2 + 4 \xi^3 \right) \Big]~,
\label{rest}
\end{eqnarray}
and the coefficients of the five elliptic integrals involved are given by
\begin{eqnarray}
f \left(\xi, c \right) & = & \frac{1}{c^2 \left( 1 + c \right)^4
  \left( c + 4 \xi^2 \right) y_1 \sqrt{\left( y_3 - y_1 \right) 
  \left(y_0 - y_2 \right)}} \nonumber \\
  & \times & \Bigg[ 4 \left( 1 + c \right) \xi^2 
  \Big( 12 \xi^3 + 2 c \, \xi \left( 1 + 6 \xi + 18 \xi^2 \right)  + 
  c^2 \left( 3 + 6 \xi + 24 \xi^2 + 68 \xi^3 \right) \nonumber \\
  & + &  2 c^3 \left( 3 + 7 \xi + 2 \xi^3 \right) \Big) + 
  4 c y_1 \Big( 10 \xi^3 \left( 1 + 2 \xi \right)  + 
  c \, \xi \left( 2 + 21 \xi + 88 \xi^2 + 120 \xi^3 + 16 \xi^4 \right) 
  \nonumber \\ & + & 2 c^2 \left( 2 + 10 \xi + 25 \xi^2 + 52 \xi^3 + 
  58 \xi^4 \right) + c^3 \left( 5 + 22 \xi + 45 \xi^2 + 64 \xi^3 + 
  64 \xi^4 + 16 \xi^5 \right) \nonumber \\
  & + & 2 c^4 \left( 1 + \xi \right)^2 
  \left( 2 + 3 \xi \right) \Big) -
  y_1^2 \left( 1 + c \right)  \left( 1 + 2 \xi \right)  
  \Big( 12 \xi^3 + 2 c \xi \left( 1 + 6 \xi + 18 \xi^2 \right) 
  \nonumber \\ & + & c^2 \left( 3 + 6 \xi + 24 \xi^2 + 68 \xi^3 \right)
  + 2 c^3 \left( 3 + 7 \xi + 2 \xi^3 \right) \Big)
  \Bigg]~,
\label{cof}
\end{eqnarray}
\begin{eqnarray}
e \left(\xi, c \right) & = & - \frac{\sqrt{(y_3 - y_1) \left(y_0 - y_2
  \right)}}{c^2 \left( 1 + c \right)^2 \left( c + 4 \xi^2 \right)}
  \Big[ 12 \xi^3 + 2 c \, \xi \left( 1 + 6 \xi + 18 \xi^2 \right) 
  \nonumber \\ & + &
  c^2 \left( 3 + 6 \xi + 24 \xi^2 + 68 \xi^3 \right) +
  2 c^3 \left( 3 + 7 \xi + 2 \xi^3 \right) \Big]~,
\label{coe}
\end{eqnarray}
\begin{eqnarray}
p \left(\xi, c \right) & = & \frac{4 \left( y_0 - y_3 \right)}{c^2 
  \left( 1 + c \right)^3 \sqrt{(y_3 - y_1) \left(y_0 - y_2 \right)}}
  \Big[ 3 \xi^2 + c \left( 2 + 4 \xi + 11 \xi^2 \right) 
  \nonumber \\ & + &
  c^2 \left( 1 + 6 \xi + 13 \xi^2 \right) + 
  c^3 \left( 2 + 6 \xi + 9 \xi^2 \right) \Big]~,
\label{cop}
\end{eqnarray}
\begin{eqnarray}
p_4 \left(\xi, c \right) & = & \frac{32 \left( y_0 - y_3 \right) 
  \xi^2 }{\left( y_5 - y_4 \right)  
  \left( 1 - 2 y_3 + y_4 - y_5 \right) \left( 1 - c + 
  \left( 1 + c \right) \left(y_5 - y_4\right)  + 4 \xi \right)^
  3 \left( c + 4 \xi^2 \right) } \nonumber \\
  & \times & \frac{1}{c^5 \sqrt{\left( y_3 - y_1 \right)  
  \left( y_0 - y_2 \right)}}
  \times \Bigg[ 12 \xi^7 + 2 c \, \xi^5 \left( 5 + 34 \xi + 
  172 \xi^2 \right) \nonumber \\ 
  & + & c^2 \xi^3 \left( 1 + 8 \xi \right) \left(2 + 37 \xi + 
  18 \xi^2 + 132 \xi^3 \right) \nonumber \\ 
  & + & c^3 \xi^2 \left( 16 + 114 \xi + 265 \xi^2 + 1002 \xi^3 + 
  212 \xi^4 + 776 \xi^5 \right) \nonumber \\
  & + &  c^4 \left( 2 + 16 \xi + 56 \xi^2 + 202 \xi^3 - 147 \xi^4 + 
  318 \xi^5 - 148 \xi^6 + 116 \xi^7 \right) \nonumber \\
  & + & c^5 \xi^2 \left( - 57 + 104 \xi - 37 \xi^2 + 68 \xi^3 - 
  24 \xi^4 \right) \nonumber \\
  & + & c^6 \xi \left(20 - 11 \xi + 2 \xi^2 - 14 \xi^3 \right) - 
  c^7 \left( 3 + 2 \xi + 2 \xi^2 \right) \nonumber \\
  & + & c \left(y_5 - y_4 \right) 
  \Big( 6 \xi^5 \left( 1 + 8 \xi \right)  + 
  c \, \xi^3 \left( 2 + 33 \xi + 86 \xi^2 + 368 \xi^3 \right) 
  \nonumber \\ & + &
  c^2 \xi^2 \left( 12 + 82 \xi + 323 \xi^2 + 158 \xi^3 + 536 \xi^4 \right) 
  \nonumber \\ & + &
  c^3 \left( 2 + 16 \xi + 56 \xi^2 + 22 \xi^3 + 313 \xi^4 - 6 \xi^5 + 
  192 \xi^6 \right) 
  \nonumber \\ & + &
  c^4 \xi^2 \left( 57 - 64 \xi + 33 \xi^2 - 52 \xi^3 + 8 \xi^4 \right) +
  c^5 \xi \left( - 20 + 5 \xi - 6 \xi^2 + 10 \xi^3 \right) 
  \nonumber \\ & + &
  c^6 \left( 3 + 2 \xi + 2 \xi^2 \right) \Big) \Bigg]~,
\label{copp4}
\end{eqnarray}
while $p_5 (\xi, c)$ can be obtained from $p_4 (\xi, c)$ by simply
interchanging $y_4$ and $y_5$.

Having obtained ${\cal F}_l (\xi, c)$, it is a simple matter to derive
the `hard' component of the characteristic function, ${\cal F}_h (\xi,
c)$ in \eq{Fy}. In fact, since the only change is in the lower limit
of integration, which for ${\cal F}_h$ coincides with one of the
branch points of the integrand, one obtains {\it the same} linear
combination of elliptic integrals, with each incomplete integral
replaced by the corresponding complete one. Furthermore, the
`remainder' rational function which would play the role of $r(\xi, c)$
in the present case vanishes. The result is
\beqa
{\cal F}_{h}(\xi,c) & = & f \left(\xi, c \right) {\bf K} 
  \Big[ m \left(\xi, c \right) \Big] 
  + e \left(\xi, c \right) {\bf E} \Big[ m \left(\xi, c \right) \Big] 
  + p \left(\xi, c \right) {\bf \Pi} \Big[ n \left(\xi, c \right), 
  m \left(\xi, c \right) \Big] \nonumber \\ & + &
  p_4 \left(\xi, c \right) {\bf \Pi} \Big[ n_4 \left(\xi, c \right), 
  m \left(\xi, c \right) \Big] + 
  p_5 \left(\xi, c \right) {\bf \Pi} \Big[ n_5 \left(\xi, c \right), 
  m \left(\xi, c \right) \Big]~.
\label{fhfin}
\eeqa
Clearly, the same technology used here can be applied to the simpler
situation in which $\xi = 0$. The characteristic function in this
limit coincides with the leading--order coefficient for the 
$C$--parameter distribution, which, to our knowledge, was never
computed in closed form before. In the limit $\xi \to 0$ one finds
\beqa
y_1 & \to & - \frac{\xi^2}{c} + {\cal O} \left( \xi^3, c^3 \right)~,
\nonumber \\
y_2 & \to & \frac{1 + 4 c - \sqrt{1 - 8 c}}{2 (1 + c)} + {\cal O} (\xi)~,
\label{y0} \\
y_3 & \to & \frac{1 + 4 c + \sqrt{1 - 8 c}}{2 (1 + c)} + {\cal O} (\xi)~,
\nonumber
\eeqa
while $y_4 \to 0$ and $y_5 \to 1$.  At this point one can get to the
result by either taking the limit $\xi \to 0$ in \eq{Fy}, and then
applying the algorithm to express that integral in terms of the basic
set of (complete) elliptic integrals, or one can take directly the
limit of \eq{fhfin}, since in the massless limit only ${\cal F}_h$
contributes. Using either method, the result takes the form
\beq
{\cal F}_0 (c) = f_0(c) \, {\bf K} 
  \Big[ m_0(c) \Big] + e_0(c) \, {\bf E} \Big[ m_0(c) \Big] 
  + p_0(c) \, {\bf \Pi} \Big[ n_0(c), m_0(c) \Big]~,
\label{f0fin}
\eeq
where, as the notation suggests, the various functions are the $\xi
\to 0$ limits of the corresponding functions in \eq{fhfin}. Some care
must be exercised in taking the limit, due to the $1/y_1$ singularity
in $f(\xi, c)$. The coefficients $p_4 (\xi, c)$ and $p_5 (\xi, c)$
vanish in the massless limit (while the corresponding integrals are
nonsingular), so that only three complete elliptic integrals appear in
the final answer. Explicitly one finds
\beqa
m_0(c) & = & \frac{2 \sqrt{1 - 8 c}}{1 - 4 c (1 + 2 c) + \sqrt{1 - 8 c}}~,
\nonumber \\
n_0(c) & = & \frac{4 \sqrt{1 - 8 c}}{(1 + \sqrt{1 - 8 \, c})^2}~,
\nonumber \\
f_0(c) & = & \frac{4 \sqrt{2} \left(1 - 2 c \, (2 + c) 
  \right)}{(1 + c)^3 \sqrt{1 - 4 c (1 + 2 c) + 
  \sqrt{1 - 8 c}}}~, \label{fun0} \\
e_0(c) & = & - \frac{3 (1 + 2 c) \sqrt{1 - 4 c (1 + 2 c) + 
  \sqrt{1 - 8 c}}}{\sqrt{2} c (1 + c)^3}~,
\nonumber \\
p_0(c) & = & \frac{\sqrt{2} (2 + c + 2 c^2)(1 - \sqrt{1 - 
  8 c})^2}{c (1 + c)^3 \sqrt{1 - 4 c (1 + 2 c) + \sqrt{1 - 8 c}}}~.
\nonumber
\eeqa
As shown in Appendix B, \eq{fun0} reproduces known results for the
leading singular behavior near $c = 0$~\cite{CW}.

\section{Expansions of the characteristic function}

Here we provide some details on the expansions of the characteristic
functions ${\cal F}_l$ and ${\cal F}_h$ at small $c$. As one can see
from \eq{Bint}, and as discussed in Section 3, the relevant scaling
limits for the computation of log--enhanced contributions to the
Sudakov exponent are $c \to 0$ with $a \equiv \xi/c$ kept constant for
${\cal F}_l$, and $c \to 0$ with $b \equiv \xi/c^2$ kept constant for
both ${\cal F}_l$ and ${\cal F}_h$.  These limits can be computed in
two different ways: either carefully taking the appropriate limit of
the integrand in \eq{Fy}, and then performing the resulting simplified
integration, or by expanding directly the elliptic integrals in
Eqs.~(\ref{flfin}) and (\ref{fhfin}). In either case one needs the
expansions of the roots of the cubic equation~(\ref{cub}). For
constant $a \equiv \xi/c$ one finds
\beqa
y_1 & = & 2 (1 - \sqrt{1 + a^2}) c - 2 a \left( 2 - a - 
  \frac{2 - a (1 - a)}{\sqrt{1 + a^2}} \right) c^2 + {\cal O} 
  \left( c^3 \right)~, \nonumber \\
y_2 & = & 2 (1 + \sqrt{1 + a^2}) c - 2 a \left(2 - a + 
  \frac{2 - a (1 - a)}{\sqrt{1 + a^2}} \right) c^2 + 
  {\cal O} \left( c^3 \right)~, \nonumber \\
y_3 & = & 1 - (1 - 2 a) c - (3 - 6 a + 4 a^2) c^2 + {\cal O} 
  \left( c^3 \right)~.
\label{coar}
\eeqa
On the other hand, for constant $b \equiv \xi/c^2$ one finds
\beqa
y_1 & = & - b^2 c^3 + {\cal O} \left( c^5 \right)~, \nonumber \\
y_2 & = & 4 c + (4 - b)^2 c^3 + {\cal O} \left( c^4 \right)~, \nonumber \\
y_3 & = & 1 - c - (3 - 2 b) c^2 - (13 - 6 b) c^3 + 
  {\cal O} \left( c^4 \right)~.
\label{cobr}
\eeqa
Similar expansions can immediately be derived for the other singular
points, $y_0$, $y_4$ and $y_5$.

Working at the level of the integrand, $f(y, c, \xi)$ in \eq{f}, the
limit $c \to 0$ with $a \equiv \xi/c$ constant is particularly
simple. In that region in fact both limits of integration behave like
constants, and one can simply expand $f(y, c, \xi)$ in powers of $c$
at fixed $a$, obtaining
\beq
f(y, c, a c) = \frac{2 (2 - (2 + a) y + y^2 )}{c y} + {\cal O}(c^0)~,
\label{fll}
\eeq
which can be immediately integrated to give the first line of
\eq{Fl_fixed_ab}.  The limit with fixed $b = \xi/c^2$ is slightly more
difficult, since in that region the lower limits of integration for
both ${\cal F}_l$ and ${\cal F}_h$ vanish linearly with $c$. Simply
expanding the integrand in powers of $c$ at fixed $b$ is not
sufficient in this case, since all powers of $c$ contribute to the
leading behavior of the integral. The reason is easily tracked to the
square--root singularity at $y = y_2$.  One can then solve the problem
by keeping exactly the factor $(y - y_2)^{-1/2}$, and expanding the
other factors of $f(y, c, \xi)$ in powers of $c$ at fixed $b$, with
the result
\beq
f(y, c, b \, c^2) = \frac{2 \left( 2 - 2y + y^2\right)}{c \sqrt{y} \sqrt{y - y_2}} + {\cal O}(c^0)~.
\label{fh}
\eeq
It is fairly easy to see that \eq{fh} gives the second line of
\eq{Fl_fixed_ab} when integrated between $b c$ and $y_3$, while it
gives \eq{Fh_fixed_b} when integrated between $y_2$ and $y_3$.

The same results were obtained by expanding directly the final
expressions for ${\cal F}_l$ and ${\cal F}_h$ in \eq{flfin} and
\eq{fhfin}, respectively.  This involves expanding the elliptic
functions in the approriate scaling limits, a non--trivial task owing
to the singularities of these functions.  One effective method to
perform this expansion is to first scale the integration variable such
that the integration limits are constants and then express the product
of square root factors as a single denominator using Feynman
parametrization.  Let us demonstrate this in the case of the
incomplete $\Pi(n,\phi,m)$ integral which contributes to the ${\cal
O}(1/c)$ terms in ${\cal F}_l$ in both limits.  Since the coefficient
multiplying this function in ${\cal F}_l(\xi,c)$ is ${\cal O}(c)$ in
these limits,
\ba
\label{p_expansion_results}
\hspace*{-30pt}
p(\xi,c) & \simeq & \left\{
\begin{array}{lr}
  \displaystyle{32(1-a)^2c}
  & \quad \quad{\rm fixed}  \,\, a  \\
   \displaystyle{32c}
   & \quad \quad{\rm fixed} \,\, b
\end{array}
\right.
\ea
relevant terms in $\Pi \left[n(\xi,c), \chi(\xi,c), m(\xi,c)\right]$
would be ${\cal O}(1/c^2)$. Starting from the definition (\ref{elli})
and defining $z\equiv \sin \phi$, one gets
\beqa
\Pi \left(n, \phi, m\right) & = &
  \frac{z}{2} \int_0^{1} \frac{d w}{\sqrt{w} \,(1 - n z^2 w) 
  \sqrt{(1 -z^2 w)(1 - mz^2 w)}}  \\ \nonumber
  & = &\frac{z}{2 \pi} \int_0^1
  \frac{d \alpha}{\sqrt{\alpha}{\sqrt{1 - \alpha}}} \int_0^{1} 
  \frac{d w}{\sqrt{w}} \,\, \frac{1}{1 - n z^2 w} \,\,\,
  \frac{1}{\alpha(1 - z^2 w) + (1 - \alpha)(1 - m z^2 w)}\,.
\eeqa
The $w$ integral can be easily done, and the result be safely
expanded in the two relevant limits.  After the expansion the
integration over the Feynman parameter~$\alpha$ can be readily
performed. The leading terms are the following:
\ba
\label{Pi_expansion_results}
\hspace*{-30pt}
\Pi \left[n(\xi,c), \chi(\xi,c), m(\xi,c)\right] & \simeq & 
  -\frac{1}{8}\frac{1}{c^2} \times\left\{
  \begin{array}{lr}
  \displaystyle{\frac{\ln{a}}{(1 - a)^2} }& \quad \quad{\rm fixed} \,\, a  
  \\  \\
  \displaystyle{\left[\ln{b c} + 2 \ln \left(\frac{1 + \sqrt{1 - 4/b}}{2}
  \right) \right]}
  & \quad \quad{\rm fixed} \,\, b
\end{array}
\right.
\ea
In the case of~${\cal F}_l$, in both limits, additional ${\cal
O}(1/c)$ contributions appear in~\eq{flfin} from $E \left[\chi(\xi,c),
m(\xi, c) \right] = 1 + {\cal O}(c^0)$, since
\ba
\label{e_expansion_results}
\hspace*{-30pt}
e \left(\xi, c \right) & \simeq & \left\{
\begin{array}{lr}
  \displaystyle{-(3 + 2 a)/{c}}
  & \quad \quad{\rm fixed}  \,\, a  \\
  \displaystyle{- 3/{c}}
  & \quad \quad{\rm fixed} \,\, b
\end{array}
\right.
\ea
Finally for fixed $a$ also $r\left(\xi, c \right)$ contributes: $r
\left(\xi, c \right)\simeq a(4+a)/c$.  Using the method described
above it is straightforward to verify that the other elliptic integral
terms in~\eq{flfin} do not contribute at order ${\cal O}(1/c)$. The
results are summarized by \eq{Fl_fixed_ab}.  The case of~${\cal F}_h$
is simpler: the elliptic integrals in~\eq{fhfin} are the complete ones
and only the fixed~$b$ limit is relevant. One gets ${\cal O}(1/c)$
contributions from ${\bf \Pi} \left[n(\xi,c), m(\xi,c)\right]\simeq
-\ln(c)/(8c^2)$ and from ${\bf E} \left[ m(\xi,c)\right]=1+{\cal
O}(c^0)$. The result is summarized by~\eq{Fh_fixed_b}.

Finally, we note that using the same techniques one can also treat the
exact expression for the leading order, which is given by the
characteristic function at $\xi = 0$ (see \eq{f0fin}).  The relevant
elliptic integrals have the asymptotic behavior
\beqa
{\bf K} \Big[m_0(c)\Big] & = & - \frac{3}{2} \log c + {\cal O} \left(
  c^3 \log c \right)~, \nonumber \\
{\bf E} \Big[m_0(c)\Big] & = & 1 + {\cal O} \left(
  c^3 \log c \right)~, \label{asell} \\
{\bf \Pi} \Big[n_0(c), m_0(c)\Big] & = & - \frac{1}{8 c^2} \log c -
  \frac{1}{4 c} \left( 1 + \log c \right) +
  {\cal O} \left( c^0 \log c \right)~, \nonumber
\eeqa
which leads to
\beq
{\cal F}_0 (c) = - \frac{3 + 4 \log c}{c} + 1 - 28 \log c +
  {\cal O} \left( c \log c \right)~,
\label{asf0}
\eeq
in agreement with Ref.~\cite{CW}, when the overall normalization is
taken into account.


\end{document}